\documentstyle[12pt]{article}
\begin{document}

\hfill{UCLA/94/TEP/25}
\hfill{UTTG-12-94}

\vspace{24pt}

\begin{center}
{\large{\bf   General Effective Actions}}

\vspace{36pt}
 Eric D'Hoker\footnote{Research supported in part by
NSF grant PHY92-18900.  E-mail address:  dhoker@physics.ucla.edu}

\vspace{4pt}
Department of Physics, University of California at Los
Angeles\\
Los Angeles, CA, 90024\\

\vspace{18pt}
Steven Weinberg\footnote{Research supported in part by the
Robert A. Welch
Foundation and NSF Grant PHY 9009850.
E-mail address: weinberg@utaphy.ph.utexas.edu}

\vspace{4pt}
Theory Group, Department of Physics, University of Texas\\
Austin, TX, 78712\\

\vspace{30pt}
{\bf Abstract}

\end{center}

\begin{minipage}{4.75in}

We investigate the structure of the most general actions with
symmetry group $G$, spontaneously broken down to a subgroup $H$.
We show that the only possible terms in the Lagrangian density
that, although not $G$-invariant, yield $G$-invariant terms in
the action, are in one to one correspondence with the generators
of the fifth cohomology classes.  For the special case of
$G=SU(N)_L \times SU(N)_R$ broken down to the diagonal subgroup
$H=SU(N)_V$, there is just one such term for $N\geq 3$, which
for $N=3$ is the original  Wess-Zumino-Witten term.

\end{minipage}

\vfill

\baselineskip=24pt
\pagebreak
\setcounter{page}{1}

Effective field theories are increasingly used to understand
the dynamics of the Goldstone bosons that result from
spontaneous breaking
of continuous symmetries.
If the action of a theory is invariant under a (compact) Lie
group $G$ of global symmetries, spontaneously broken to a
subgroup $H$, then the Goldstone
fields $\pi ^a (x)$ in the effective action parametrize the
coset space
$G/H$ with $a=1, \cdots, {\rm dim}~ G/H$, and accordingly
transform under linear representations of $H$, but under
non-linear realizations of the broken symmetries of $G$.
The power of effective field theories arises largely from
the fact that the nonlinearly realized broken symmetry
allows only a finite number of terms in the action, up to
any given order in an expansion in powers of derivatives or
momenta.

A general method for constructing invariant non-linear
effective actions was given in Ref. [1] for $SU(2)_L \times
SU(2)_R$
and was extended to the case of arbitrary $G$ and $H$ in
Ref. [2].
But although this method yields the most general
$G$-invariant term in the effective Lagrangian, its results are
not quite complete.  Wess and Zumino [3] showed that fermion
loops produce a four-derivative term in the effective
Lagrangian for the strong-interaction Goldstone octet that
is not invariant under $SU(3)\times SU(3)$, but rather
changes under $SU(3)\times SU(3)$ transformations by a
total derivative, so that the action {\it is} $SU(3)\times
SU(3)$ invariant.    Subsequently Witten [4] was able to
re-express this term as the integral over an invariant
Lagrangian density in five dimensions.  The WZW action has
since then been generalized in Ref. [5] to $G/H$ models with
arbitrary $G$ and $H$.

It is natural to ask whether there are any more possible
terms in the action (not necessarily related to anomalies in
the underlying theory),  that, although invariant under a
nonlinearly realized symmetry $G$, are not the
four-dimensional integrals of $G$-invariant Lagrangian densities.
This question seems to us important, as the effective field
theory approach is based on our ability to catalog {\it all}
invariant terms in the action with a given number of
derivatives.

The first step is to show that even where the action is not
the integral of a $G$-invariant Lagrangian density, its
variation with respect to the Goldstone boson fields is an
invariant density.  The Goldstone boson fields $\pi^a(x)$
enter the action as a parameterization of a general
spacetime-dependent $G$-transformation $U(\pi(x))$, so the
variation of the action under an arbitrary change in $\pi$
may be written as
\begin{equation}
\delta S[\pi]=\int d^4x\;{\rm Tr}\,\left\{(U^{-1}\delta
U)_{\cal X}\,J\right\}\;,
\end{equation}
where a subscript ${\cal X}$ or ${\cal H}$ will denote the
terms proportional to the broken and unbroken symmetry
generators $x_a$ and $t_i$, respectively, and the
coefficient $J$ is a local function of the Goldstone boson
fields and their derivatives.  Let us work out how $J$
transforms.  According to the general formalism of [2],
under a global transformation $g\in G$, the Goldstone boson
fields undergo the transformation $\pi\rightarrow \pi'$,
with
\begin{equation}
g\,U(\pi)=U(\pi')\,h(\pi,g)\;,
\end{equation}
where $h(\pi,g)$ is some element of the unbroken subgroup
$H$.
Since $S[\pi]=S[\pi']$ for all $\pi$, the variational
derivatives are also equal
$$\frac{\delta S[\pi']}{\delta \pi^a}=\frac{\delta
S[\pi]}{\delta \pi^a}\;.
$$
(Note that the derivative is with respect to $\pi$, not
$\pi'$, on both sides of the equation.)  Using Eq. (1), this
is
\begin{equation}
{\rm Tr}\,\left\{\left[U^{-1}(\pi')\frac{\partial
U(\pi')}{\partial \pi^a}\right]_{\cal
X}\,J(\pi')\right\}={\rm Tr}\,\left\{\left[U^{-
1}(\pi)\frac{\partial U(\pi)}{\partial \pi^a}\right]_{\cal
X}\,J(\pi)\right\}\;.
\end{equation}
To put this in a useful form, take the derivative of Eq (2)
with respect to $\pi^a$, and multiply on the left with
$U(\pi')^{-1}$ and on the right with $h^{-1}(\pi,g)$:
$$
U^{-1}(\pi')\frac{\partial U(\pi')}{\partial \pi^a}=
h(\pi,g)\,U^{-1}(\pi)\frac{\partial U(\pi)}{\partial
\pi^a}\,h^{-1}(\pi,g)-\frac{\partial h(\pi,g)}{\partial
\pi^a}\,h^{-1}(\pi,g)
$$
and so
\begin{equation}
\left[U^{-1}(\pi')\frac{\partial U(\pi')}{\partial
\pi^a}\right]_{\cal X}=
h(\pi,g)\,\left[U^{-1}(\pi)\frac{\partial U(\pi)}{\partial
\pi^a}\right]_{\cal X}\,h^{-1}(\pi,g)\;.
\end{equation}
Eq. (3) then becomes
\begin{equation}
{\rm Tr}\,\left\{\left[U^{-1}(\pi)\frac{\partial
U(\pi)}{\partial \pi^a}\right]_{\cal X}\,\left[h^{-
1}(\pi,g)\,J(\pi')\,h(\pi,g)\right]\right\}={\rm
Tr}\,\left\{\left[U^{-1}(\pi)\frac{\partial U(\pi)}{\partial
\pi^a}\right]_{\cal X}\,J(\pi)\right\}\;.
\end{equation}
 From linear combinations of the quantities $[U^{-
1}(\pi)\,\partial U(\pi)/\partial \pi^a]_{\cal X}$ we can
form arbitrary linear combinations [6]
 of the broken symmetry generators $x_a$, so (5) yields the
transformation rule for $J$:
\begin{equation}
J(\pi')=h(\pi,g)\,J(\pi)\,h^{-1}(\pi,g)\;.
\end{equation}
  Following the same
arguments that led to (4), we easily see that also
\begin{equation}
\left[U^{-1}(\pi')\delta U(\pi')\right]_{\cal X}=
h(\pi,g)\,\left[U^{-1}(\pi)\delta U(\pi)\right]_{\cal
X}\,h^{-1}(\pi,g)\;,
\end{equation}
so ${\rm Tr}\,\{(U^{-1}\delta U)_{\cal X}\,J\}$ is invariant
under $G$.

 This result leads to a natural five-dimensional formulation
of the theory.  As usual, we compactify spacetime to a
four-sphere $M_4$ by requiring that all fields  approach
definite limits as $x^\mu\rightarrow\infty$.  The operator
$U(\pi(x))$ therefore traces out a four-sphere in the
manifold of $G/H$ as $x^\mu$ varies over $M_4$.  If the
homotopy group $\pi_4(G/H)$ is trivial (as is the case for
$SU(N)\times SU(N)$ spontaneously broken to $SU(N)$ with
$N\geq 3$), or if $U(\pi(x))$ belongs to the trivial element
of $\pi_4(G/H)$, then we may introduce a smooth function
$\tilde{\pi}^a(x,t_1)$, such that
$\tilde{\pi}^a(x,1)=\pi^a(x)$, and $\tilde{\pi}^a(x,0)=0$.
In this way spacetime is extended to a five-ball $B_5$ with
boundary $M_4$ and coordinates $x^\mu$ and $t_1$.  The
action may then be written in the five-dimensional form
\begin{equation}
S[\pi]=\int_{B_5}d^4x\, dt_1\; {\cal L}_1\;,
\end{equation}
where ${\cal L}_1$ is the $G$-invariant density ${\rm
Tr}\,\left\{(U^{-1}\,\partial U/\partial t_1)_{\cal X}
J\right\}$.
(When $\pi _4 (G/H)\not=0$, we may interpolate
between $\pi ^a (x)$
and a fixed representative $\pi _o ^a(x)$ of the homotopy class
of $\pi ^a (x)$.
The difference $S[\pi]-S[\pi _o]$ is given by the integral
over the cylinder $M_4 \times [0,1]$ of the same density
${\cal L}_1$ as in (8) and the arguments
to be presented below still hold. In some cases, $G/H$ may
be naturally embedded into a larger space with vanishing
fourth homotopy group, as is the case for $SU(2)$ embedded
in $SU(3)$, considered in [4].)

We next show that this is the integral of a $G$-invariant
5-form on $G/H$.  Consider a general deformation
$\pi(x)\rightarrow \tilde{\pi}(x;t)$, where $t^i$ are a set
of dim$(G/H)-4$ free parameters, that along with the $x^\mu$
provide a set of coordinates for $G/H$.  The coordinate
$t_1$ in (8) can be chosen to be any one of these
parameters.  We have shown that
\begin{equation}
\frac{\partial S[\tilde{\pi}]}{\partial t_i}=\int_{M_4} d^4
x\;{\cal L}_i\;,
\end{equation}
where ${\cal L}_i\equiv {\rm Tr}\,\{(\tilde{U}^{-1}\partial
\tilde{U}/\partial t^i)_{\cal X}\,J\}$
are $G$-invariant functions of $\tilde{\pi}^a$ and its
derivatives.  The general rules of [2] would allow a wide
variety of terms in ${\cal L}_i$, but these are limited by
integrability conditions.  From (9) we see that
$$\int_{M_4} d^4 x\;\left(\frac{\partial{\cal L}_i
}{\partial t^j}
-\frac{\partial{\cal L}_j }{\partial t^i}\right)=0\;.
$$
Since this integral vanishes for all $\tilde{\pi}(x)$, its
integrand must be an $x$-derivative [7]:
$$\frac{\partial{\cal L}_i }{\partial t^j}
-\frac{\partial{\cal L}_j }{\partial t^i}=-\partial_\mu
{\cal L}^\mu_{ij}\;.
$$
This can be written in the language of differential forms,
as $d_t F_1=-d_x F_2$, where
$$d_t\equiv dt^i\partial_i~~~~~~~~~~~d_x\equiv
dx^\mu\partial_\mu$$
and $F_1$ and $F_2$ are the five-forms
$$
F_1\equiv\frac{1}{24} \epsilon_{\mu\nu\rho\sigma}{\cal
L}_i\, dt^idx^\mu dx^\nu dx^\rho dx^\sigma
{}~~~~~~~~~~~
F_2\equiv \frac{1}{12}\epsilon_{\mu\nu\rho\sigma}{\cal
L}_{ij}^\mu \,dt^i dt^j
  dx^\nu dx^\rho dx^\sigma \;.$$
It follows that $0=d_t^2 F_1=d_x(d_t F_2)$, so by an
extension of Poincar\'e's Lemma, in any simply connected
patch we will have $d_t F_2= -d_x F_3$, where $F_3$ is a
five-form $\epsilon_{\mu\nu\rho\sigma}{\cal
L}_{ijk}^{\mu\nu}dt^i dt^j dt^k dx^\mu dx^\nu$.  Continuing
in this way, we can construct five-forms $F_4$ and $F_5$
proportional respectively to four and five $dt$ factors,
with $d_tF_3=-d_xF_4$, $d_tF_4=-d_xF_5$, and $d_tF_5=0$.
Hence $F\equiv \sum_{N=1}^5 F_N$ is a closed five-form on
$G/H$:
\begin{equation}
dF=0~~~~~~~~~~~~~~~~~~~~d\equiv d_x+d_t\;.
\end{equation}
Also, because $B_5$ has $t_2, t_3$, etc., all constant, Eq.
(8) may be written
\begin{equation}
S[\pi]=\int_{B_5} F\;.
\end{equation}

So far, only the term $F_1$ has been shown to be
$G$-invariant.  The group $G$ acts transitively on the manifold
$G/H$, so a $G$ transform of a form is always continuously
connected to the original form.  Thus the two forms are
homotopic and define the same de Rham cohomology class.  One
can construct a $G$-invariant form in this cohomology class
by integrating the form over the group $G$ with the
invariant Haar measure [8,9].   This has no effect on (11), since
the integral depends only on $F_1$, which is already
invariant.  Also, one can similarly show that any two
invariant p-forms in the same cohomology class differ not
only by an exterior derivative, but specifically by the
exterior derivative of an {\it invariant} $(p-1)$-form.
Such an exterior derivative term in the five-form $F$ would
yield a term in (8) that can be written as the
four-dimensional integral of a $G$ invariant density, so the
classification of terms in $S[\pi]$ that {\it cannot} be so
written is now reduced to the problem of finding the fifth
de Rham cohomology group  $H^5(G/H;{\bf R})$ of the manifold
$G/H$ [10].

The fifth de Rham cohomology group is well known where $G/H$
is itself a simple Lie group.  For $G=SU(N)$ with $ N\geq 3$
(including the case $SO(6)\sim SU(4)$), $H^5(G;{\bf R})$
has a single generator
\begin{equation}
\Omega_5=\frac{-i}{ 240 \pi ^2}{\rm Tr}\, (U^{-1}d~U)^5\;.
\end{equation}
(Here and henceforth, we suppress wedges in the exterior
product of differential forms, reserving them for the products
of cohomology groups.)
This is in particular the case for $SU(N)\times SU(N)$
spontaneously broken to $SU(N)$ with $N\geq 3$, where $G/H$
is itself just $SU(N)$.  Eq. (12) is the original
Wess-Zumino-Witten
term, which we now see is indeed unique.  All
other simple (or $U(1)$) Lie groups  have trivial fifth
cohomology groups.  For the original case [1] of
$SU(2)\times SU(2)$ spontaneously broken to $SU(2)$ the
cohomology is trivial, so all invariant actions are the
integrals of invariant Lagrangian densities.

Where $G/H$ is a product space, we use the
K\"unneth formula [8,9]:
\begin{equation}
H^k (K_1 \times K_2;{\bf R})
= \sum _{k_1+k_2=k} H^{k_1}(K_1;{\bf R}) \wedge H^{k_2}
(K_2;{\bf R})\;,
\end{equation}
which gives $H^5(G/H;{\bf R})$ in terms of the cohomologies of
its factors up to degree 5.
For this purpose, we  need to know that [11,12,13]  for all
simple Lie groups $G$, $H^k(G;{\bf R})$ vanishes for $k=1,2,4$
while $H^3(G;{\bf R})$ has a single generator
(corresponding to the Goldstone-Wilczek topologically
conserved current [14] )
\begin{equation}
\Omega _3 = { i \over 12 \pi} {\rm Tr}\, (U^{-1}d~U)^3\;.
\end{equation}
Also $H^k(U(1);{\bf R})$ vanishes for $k>1$, while for $k=1$ it
has a single generator
\begin{equation}
\Omega _1 = -i{\rm Tr}\, (U^{-1}d~U)\;.
\end{equation}
Finally, $H^o(K;{\bf R})={\bf R} ^c$, where $c$ is the number of
connected components of $K$; for our purposes this just
means that if $H^5(K;{\bf R})$ for some space $K$ has a generator
$\Omega_5$, then $H^5(K\times K';{\bf R})$ has the same generator
for any $K'$.
To each generator of $H^5(G/H;{\bf R})$, there corresponds a
WZW-like  term in the five-dimensional Lagrangian, and an
independent
coupling constant.  In particular, if $G$ is semi-simple,
with precisely $p$ factors $SU(N_i)$ with $N_i\geq 3$ and
all other simple factors with $H^5=0$, then we have $p$
different terms of the Wess-Zumino-Witten type, each of
which has an independent coupling constant in the action.
This result is of course expected for a product of groups,
and is known to appear explicitly  in the low energy
effective action when massive fermions are integrated out of
the path integral [15].

When $G/H$ is not itself a Lie group, the
fifth cohomology group of $G/H$ may still be obtained from that of
$G$.  For any simple group $G$ and subgroup $H$, we may
construct a `projected' five-form on $G/H$ that is invariant
under local $H$
transformations [5,15,16,17], and is given by:
\begin{equation}
\Omega _5 (U;V)
 = \frac{-i}{ 240 \pi ^2} \bigl \{
	{\rm Tr}\, (U^{-1} D U)^5 - 5{\rm Tr}\, W (U^{-1} D U)^3
        +10 {\rm Tr}\, W^2 U^{-1} D U \bigr \}\;,
\end{equation}
where $V$ is the $H$-connection $V=(U^{-1}d~U)_{\cal H}$,  $DU$ is
the $H$-covariant derivative $DU=dU-UV$, and the trace is
evaluated in any convenient representation of $G$, usually
taken as the defining representation.    In general, $\Omega
_5 (U;V)$ is neither closed nor simply related to the
generator $\Omega_5(U;0)$ of $H^5(G;{\bf R})$.  Rather,
\begin{equation}
d \Omega _5 (U;V)
 =
\frac{i}{ 24 \pi ^2}~d_{rst} ~W^r  W^s  W^t
\end{equation}
and
\begin{equation}
\Omega _5 (U;V)=\Omega _5 (U;0) + \Omega _5 (1;V) +d~\gamma
(U;V)\;,
\end{equation}
where $W$ is the field strength $W=dV+V^2  $, and
$d_{rst}$ is the trace of the symmetrized product of
generators $\rho^r$ of $H$, $ 2d_{rst} = {\rm Tr}\,  \rho^r
\{\rho^s, \rho^t \} $,  which plays a key role in the study
of the chiral anomaly in four dimensions [18].  But if
$d_{rst}=0$, then the five form $\Omega_5(U;V)$ is closed,
and also each term in the 5-dimensional Chern-Simons term
$\Omega _5(1;V)$
for the ${\cal H}$-valued gauge field $V$ vanishes.
The form $\Omega _5(U;V)$ then belongs to the same
cohomology
class as $\Omega _5(U;0) $ and it can be shown that
there is a one to one correspondence
between the fifth cohomology generators of $G/H$ and those of $G$.
On the other hand, if $d_{rst}\neq 0$, then the projected
form of (16) is not closed and it can be shown that the
fifth cohomology is trivial in this case.
For example, any coset space of the type $SU(n)/H$ with
$n\geq 3$,
where $H$ is embedded in $G$ in such a way that
$d_{rst}=0$, has one cohomology generator, given in (16).

It is noteworthy that the simple groups $SU(2);
{}~Sp(2N);~SO(N), ~N\geq 7; ~E_6,~E_7,~E_8;~F_4,~G_2$ that
have zero fifth cohomology are also those that have
vanishing
$d$ symbols.  We now see that for such groups, the coset
spaces $G/H$ have $H^5(G/H;{\bf R})~=0$ for all subgroups $H$.
These properties are easily verified for
the special case of compact symmetric spaces [12,13].
Also, when rank$(G)$=rank$(H)$,
a classic theorem [13]
states that all odd cohomology classes vanish.  An example
of a general class of manifolds $G/H$ with rank$(H)\leq $
rank$(G)$ for which $H^5(G;{\bf R})~ \not=0$ and
 $H^5(G/H;{\bf R})=0$  is provided [12] by
$$
SU(n)/S(U(k_1) \times \cdots \times U(k_q)) \qquad k=\sum
_{\alpha =1}
^q k_\alpha\geq 3\;,
$$
with the $U(k_\alpha)$ embedded in $SU(n)$ in such a way
that the defining representation of $SU(n)$ transforms also
as the defining representation of $S(U(k_1) \times \cdots
\times U(k_q))$.

Finally, if $G$ is not simple, and $H$ is a non-trivial
subgroup,
the cohomology problem can be solved by analyzing the
restriction
of the $d$-symbols of $G$ to the subgroup $H$ [19].
If $G$ is semi-simple (and $H$ is connected),
two types of cohomology generators arise [20].
First, the projected form of (16) is now obtained as a linear
combination of $\Omega _5 (U;V)$ on each simple component
of $G$ with non-vanishing fifth cohomology.
Linear combinations for which $d_{rst}$ on the subgroup $H$
vanishes, yield generators of $H^5(G/H;R)$.
Second, there may be generators that are linear combinations
of products of cohomology generators on $G/H$ of degrees 2 and 3.
Generators of degree 2 correspond to the field strength
associated with generators of invariant Abelian subgroups of $H$
(i.e. U(1) factors). Generators of degree 3 correspond to
the Goldstone-Wilczek current of (14), projected to $G/H$.
When $G$ is not semi-simple and contains
extra U(1) factors, there are also linear combinations of products
of generators of degree 1 with generators of degrees 1, 2, 3 and 4.

We conclude with a brief discussion of global quantization
conditions. Different interpolating maps are generally
topologically
inequivalent (their equivalence classes being given by
$\pi _5 (G/H)$), and there is no natural way of choosing one
interpolation
above another. Witten has argued that the quantum action can
be
allowed to be multiple valued, provided the action changes
additively by
integer multiples of $2\pi$.[4] The dependence of interpolation
becomes invisible in the quantum theory provided the
coupling
constants multiplying $\Omega _5$ as normalized in (12) are
integers.
In the present case, this quantization condition must be
enforced
on every independent coupling constant multiplying each
non-trivial WZW term normalized as in (12).

A slight refinement of this quantization condition is
required
when $\pi _4 (G)=0$ and $\pi _4(H) \not=0$.
For all simple groups $H$ we have $\pi _4(H)=0$, except
when $H$ is a symplectic group, for which
$\pi _4 (Sp(2n))={\bf Z}_2$.
Whenever $\pi_4(H)={\bf
Z}_2$, $H$ has a discrete anomaly [21], even though its
$d$-symbols vanish identically, and it can be shown that the
coupling constant of the corresponding term of $H^5(G/H;{\bf
R})$ must be quantized in terms of {\em even} integers to
obtain a single-valued path integral [22].

We are glad to acknowledge several valuable conversations
with E. Farhi, who collaborated in the early stages of this
work.  We benefited from helpful conversations of one of us
(E. D'H.) on cohomology with V. S. Varadarajan, and of the
other (S.W.) with E. Witten. We thank H. Leutwyler for drawing
our attention to his related work in [23].

\bigskip
\bigskip

\noindent
{\bf References}

\begin{enumerate}

\item S. Weinberg, Phys. Rev. {\bf 166},  1568 (1968).

\item S. Coleman, J. Wess and B. Zumino, Phys. Rev. {\bf 177},
2239 (1969) ; C.G. Callan, S. Coleman, J. Wess and B. Zumino,
Phys. Rev. {\bf 177}, 2247 (1969).

\item J. Wess and B. Zumino, Phys. Lett.  {\bf 37B},  95
(1971).
\item E. Witten,  Nucl. Phys. {\bf B223}, 422 (1983).

\item Y.-S. Wu, Phys. Lett. {\bf 153B},  70
(1985); C.M. Hull and B. Spence, Nucl. Phys. B353 (1991) 379.

\item With the
exponential parameterization,  $U^{-1}(\pi)\,\partial
U(\pi)/\partial \pi^a \rightarrow ix_a$ for small $\pi$, so
the quantities  $[U^{-1}(\pi)\,\partial U(\pi)/\partial
\pi^a]_{\cal X}$ span the same space as the $x_a$ for $\pi$
in at least a finite neighborhood of the origin.

\item Here we are
using a general theorem, that if the integral over a closed
manifold of a local function of a field and its derivatives
vanishes for all such fields, then the integrand must be a
derivative of another local function of the field and field
derivatives.  Since we do not know where this theorem is to
be found in the mathematical literature, we have proven it
by direct construction of the latter function.

\item S. Kobayashi and K. Nomizu, {\it Foundations of
Differential
        Geometry}, Vol II (J. Wiley \& Sons, 1969).

\item B.A. Dubrovin, A.T. Fomenko and S.P. Novikov, {\it
        Modern Geometry and Applications}, Vol III (Springer
Verlag, 1990).

\item The above arguments
may easily be carried over to the construction of invariant
actions in space-times of dimension $d$,
where the allowed non-invariant Lagrangian densities are in one to one
correspondence with the generators of $H^{d+1}(G/H;{\bf R})$.

\item {\it Encyclopedic Dictionary of Mathematics},
                 S. Iyanaga and Y. Kawada, eds. (MIT Press,
1980).

\item W. Greub, S. Halperin and R. Vanstone, {\it
Connections,
        Curvature and Cohomology}, Vol III, (Acad. Press,
1976).

\item A. Borel, Ann. Math. (2) {\bf 57}, 115 (1953); also in
A. Borel, {\it Collected Papers}, Vol I (Springer Verlag,
1983).

\item J. Goldstone and F. Wilczek, Phys. Rev. Lett. {\bf 47}, 986
(1981).

\item E. D'Hoker and E. Farhi, Nucl. Phys. {\bf B248}, 59,
77  (1984).

\item S.S. Chern, {\it Complex Manifolds without Potential
Theory}  (Springer Verlag, 1979).

\item B. Zumino, in {\it Relativity, groups and Topology II: Les Houches 1983},
B. De Witt, R. Stora, eds. (North Holland, 1984);
K. Chou, H.Y. Guo, K. Wu and X. Song, Phys. Lett. {\bf 134B},  67 (1984);
J. Manes, R. Stora and B. Zumino, Comm. Math. Phys. {\bf 102},
157 (1985); J. Manes, Nucl. Phys. {\bf B250}, 369
 (1985).

\item W.A. Bardeen, Phys. Rev. {\bf 184}, 1848  (1969);
D.J. Gross and R. Jackiw, Phys. Rev. {\bf D6}, 477  (1972);
H. Georgi, {\it Lie Algebras in Particle Physics}, (Benjamin/Cummings, 1982).

\item H. Cartan, in {\it Colloque de Topologie, Centre Belge
de Recherches Math\'e-matiques, Brussels 1950}, (G. Thone, 1950).

\item A discussion of these results will be presented elsewhere.

\item E. Witten, Phys. Lett. {\bf 117B}, 324  (1982).

\item This includes the cases
$SU(2)\sim SO(3)\sim Sp(2)$ and $SO(5)=Sp(4)$.

\item H. Leutwyler, ``Foundations of Chiral Perturbation Theory",
Univ. Bern preprint, BUTP-93/24, to appear in Annals of Physics.

\end{enumerate}
\end{document}